\documentclass{kapproc} 

\usepackage{procps} 

\usepackage[dvips]{graphicx}

\upperandlowercase

\kluwerbib 

\begin{document}

\articletitle[The Hierarchical Formation of the Galactic Disk]
{The Hierachical Formation of the \\ Galactic Disk}

\author{Julio F. Navarro}
 
\affil{CIAR and Guggenheim Fellow, Department of Physics and Astronomy, University of Victoria, Victoria, BC, Canada.}

\begin{abstract}
I review the results of recent cosmological simulations of galaxy
formation that highlight the importance of satellite accretion in the
formation of galactic disks. Tidal debris of disrupted satellites may
contribute to the disk component if they are compact enough to survive
the decay and circularization of the orbit as dynamical friction
brings the satellite into the disk plane. This process may add a small
but non-negligible fraction of stars to the thin and thick disks, and
reconcile the presence of very old stars with the protracted
merging history expected in a hierarchically clustering universe. I
discuss various lines of evidence which suggest that this process may
have been important during the formation of the Galactic disk.
\end{abstract}


\section{Introduction}

Stellar disks are traditionally viewed as ensembles of stars on nearly
circular orbits, dynamically fragile to fluctuations in the
gravitational potential brought about by accretion events and
mergers. Indeed, it is common to use the oldest stars in the Milky Way
disk to date the time of its last major merger. Indications that star
formation in the solar neighbourhood has been ongoing for most of the
age of the Universe, coupled with the presence of metal-poor, old
stars on disk-like orbits in the vicinity of the Sun, suggest that the
Galaxy has suffered few, if any, important accretion events in the
past $\sim 10$ Gyr. This relatively peaceful evolution is
at odds with the rather hectic merging activity expected in
hierarchically clustering structure formation models (see, e.g.,
Steinmetz \& Navarro 2000).

Recent numerical simulations (Abadi et al 2003a,b) suggest one way of
reconciling the flurry of early mergers expected in a $\Lambda$CDM
universe---the current paradigm of structure formation models---with
the presence of an old disk component in the Milky Way. These
simulations show that satellite accretion events may contribute stars
not only to the spheroid, but also to the disk component of a
galaxy. These stars make up the core of satellites whose orbits are
eroded, circularized, and brought into the plane of the galaxy before
being disrupted. Such events contribute upwards of $\sim 10\%$ of
stars in the disk, and the majority of old disk stars.

In this contribution I describe several lines of evidence that suggest
that, indeed, the Milky Way disk may contain a number of stars which
originate in the tidal debris of disrupted satellites. If confirmed by
further scrutiny, this evidence would demonstrate that mergers and
accretion events have been responsible for shaping the disk of the
Milky Way, as well as its spheroid, and would provide strong support
to the hierarchical mode of galaxy assembly envisioned in the
$\Lambda$CDM scenario.

\section[]
{Tidal debris in the disk of the Milky Way}

Since the bulk of the thin disk of the Galaxy is made up of stars that
formed {\it in situ} following the collapse of a gaseous, dissipative
component, left-over debris from past accretion events is most easily
identified in samples of stars that minimize contamination by the thin
disk. Thus, tidal debris should show more prominently in
metal-deficient star samples, since these are likely to contain stars
that formed before the merging activity abated, as well as in samples
collected above, below, or in the outskirts of, the Galactic plane.

One prime example of this is the discovery by the SDSS team of a
``ring-like'' structure in the direction of the anti-galactic center
(Yanny et al 2003). Further studies have confirmed that this is a
dynamically coherent structure that spans a large arc on the plane of
the Galaxy located at roughly 18 kpc from the Galactic
center. Numerical simulations show that these ``tidal arcs'' occur
naturally during the disruption of satellites on orbits coplanar with
the disk (Helmi et al 2003), and have identified features that
distinguish this accretion interpretation from other competing
scenarios, such as spiral arms or the resonant response of the disk to
the influence of the Magellanic Clouds or of the bar in the Galactic
bulge.

One example of ``tidal arcs'' found in the simulation of Abadi et al
(2003a,b) is shown in Figure 1. (Details may be found in Helmi et al
2003.) The arc circumscribed azimuthally between $\phi=100^o$ and
$\phi=210^o$ is caused by the apocentric ``wrapping'' of the inner
tidal arm stripped from the satellite after a recent pericentric
passage. The arc is fed by a continuous stream of particles escaping
the satellite. There is a clear gradient in the energies of particles
in the arc; the most bound are most advanced along the arc
(i.e. larger $\phi$), which is reflected in a clear velocity gradient
across the arc, shown in the middle panel of Figure 1.

This arc is a short-lived transient feature that weakens as particles
of different energy phase mix throughout the disk. Thus, if this
interpretation is correct, one the SDSS ``ring'' to be dynamically
``young'', and that the parent (disrupting) satellite may be still
lurking somewhere in the Galactic disk.  Recently, there have been
intriguing suggestions that oddities in the distribution of disk stars
in the direction of Canis Major may be due to the core of a disrupting
dwarf rather than to the presence of a warp in the outer Galactic disk
(Martin et al 2004). 
I conclude that the
evidence in favour of interpreting the SDSS ``ring'' as debris from a
(probably ongoing) accretion event is, if not conclusive, at least
very compelling.

\begin{figure}[t]
\vskip.2in
\centerline{\includegraphics[width=5in]{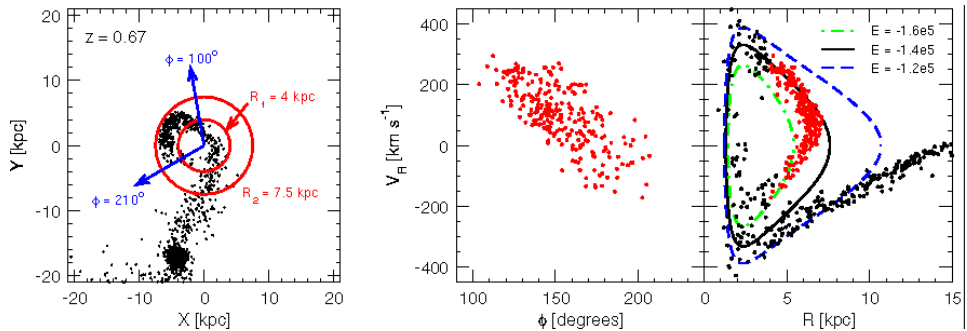}}
\caption{Snapshot of the disruption of a satellite on an orbit roughly
coplanar to the disk, taken from the simulation of Abadi et
al. (2003a,b).  Left panel shows a ``face on'' projection of the
debris shortly after the first pericentric passage. The main galaxy is
at the center of coordinates, but only the stars in the satellite are
shown for clarity. Middle panel shows, in a cylindrical coordinate
system, the (galactocentric) radial velocity versus azimuthal angle
$\phi$ for particles in the arc.  Right hand panel shows radial
velocity and distance for all satellite particles. Curves in this
panel indicate the loci of particles with three different values of
the binding energy, $E$, and angular momentum equal to the average of
particles in the arc. See Helmi et al (2003) for details.}
\end{figure}

Are there other examples of this process? We have begun to scrutinize
dynamically coherent structures in samples of stars that favour
metal-deficient stars, in particular the compilations of Beers et al
(2000, B00) and Gratton et al (2003, GCCLB). Figure 2a shows the
distribution of specific angular momentum of stars in the B00 and
GCCLB samples. These compilations favour metal-poor stars, and
therefore contain mainly stars in the solar neighbourhood that belong
to the traditional spheroid and thick disk components. As one can see
in Figure 2a, the $J_z$-distribution is not smooth, and is
punctuated by a number of local ``peaks'', even when only stars with
orbits confined to the plane are considered.

These ensembles of stars with common rotation speeds are reminiscent
of the dynamical substructures that Eggen (1986 and references
therein) identified as ``moving groups'' in the solar
neighbourhood. Eggen proposed that these groups were the late stages
in the disruption of open clusters, recognizable locally as ensembles
of stars with negligible dispersion in their rotation
speed. Unfortunately, the lack of accurate distances for most
candidates complicated the analysis and prevented proper assessment of
his hypothesis. We (Navarro, Helmi \& Freeman 2003) have reanalyzed
Eggen's hypothesis for the case of the ``Arcturus group''.

Arcturus, the fourth brightest star in the night sky, is a Population
II giant whose odd kinematics has been documented for centuries. It is
very close to the Sun (hence its apparent brightness), and moves on an
orbit confined to the Galactic plane, yet its rotation speed is only
{\it half} that of the Sun. The vertical line labelled ``Arcturus'' in
Figure 2a shows that there is a significant excess (over a smooth
distribution) of stars of similar angular momentum as Arcturus in the
B00 and GCCLB compilations. 

Although there is a peak in the $J_z$ distribution of stars in the B00
compilation that coincides with Arcturus, further inspection shows
that these stars span a wide range of metallicities, casting doubt on
Eggen's interpretation of this group as a disrupting star
cluster. Furthermore, the dispersion in rotation speed of the group
far exceeds the $\sim 0.5$ km s$^{-1}$ that would be characteristic of
a disrupting cluster. The metal abundances of stars in the group,
however, show a distinct pattern, as shown in Figure 2b. This figure
shows the enrichment in $\alpha$ elements ([$\alpha$/Fe]) as a
function of iron abundance, in solar units ([Fe/H]), for all stars in
the GCCLB compilation (open circles). ($\alpha$ element abundances are
not available for stars in the B00 catalog.)

The solid circles in Figure 2b correspond to those stars in the bin
labelled ``Arcturus'' in Figure 2a. The trend defined by these stars
is distinct from that traced by all GCCLB stars. The Arcturus group
candidates define a narrow path in the [$\alpha$/Fe] vs [Fe/H] plane
well approximated by one of the simple closed-box self-enrichment
models of Matteucci \& Francois (1989, see solid line in Figure
2b). Thus, GCCLB stars with $J_z$ comparable to Arcturus form a
dynamically and chemically coherent group of stars which might be
naturally identified with debris from a disrupted satellite
galaxy. Further corroborating evidence is provided by the discovery by
Gilmore, Wyse \& Norris (2002) of a population of stars above and
below the plane of the Galaxy with angular momentum similar to
Arcturus, which they interpret as the debris of a tidally disrupted
satellite. 

The discovery of evidence for past (and perhaps ongoing) accretion
onto the disk of the Milky Way suggests a reassessment of the
traditional scenarios for the assembly and enrichment of stars in the
solar neighbourhood. Extrapolating boldly, one may even argue that
{\it most} metal-deficient stars (disk and spheroid) in the solar
neighbourhood have been contributed by various accretion events
throughout the life of the Galaxy. Confirming such assertion would
clarify the role of mergers and accretion events in the formation of
the Galactic disk and would lend strong support to hierarchical
theories of galaxy formation.

\begin{figure*}[t]
\includegraphics[height=0.32\textheight,clip]{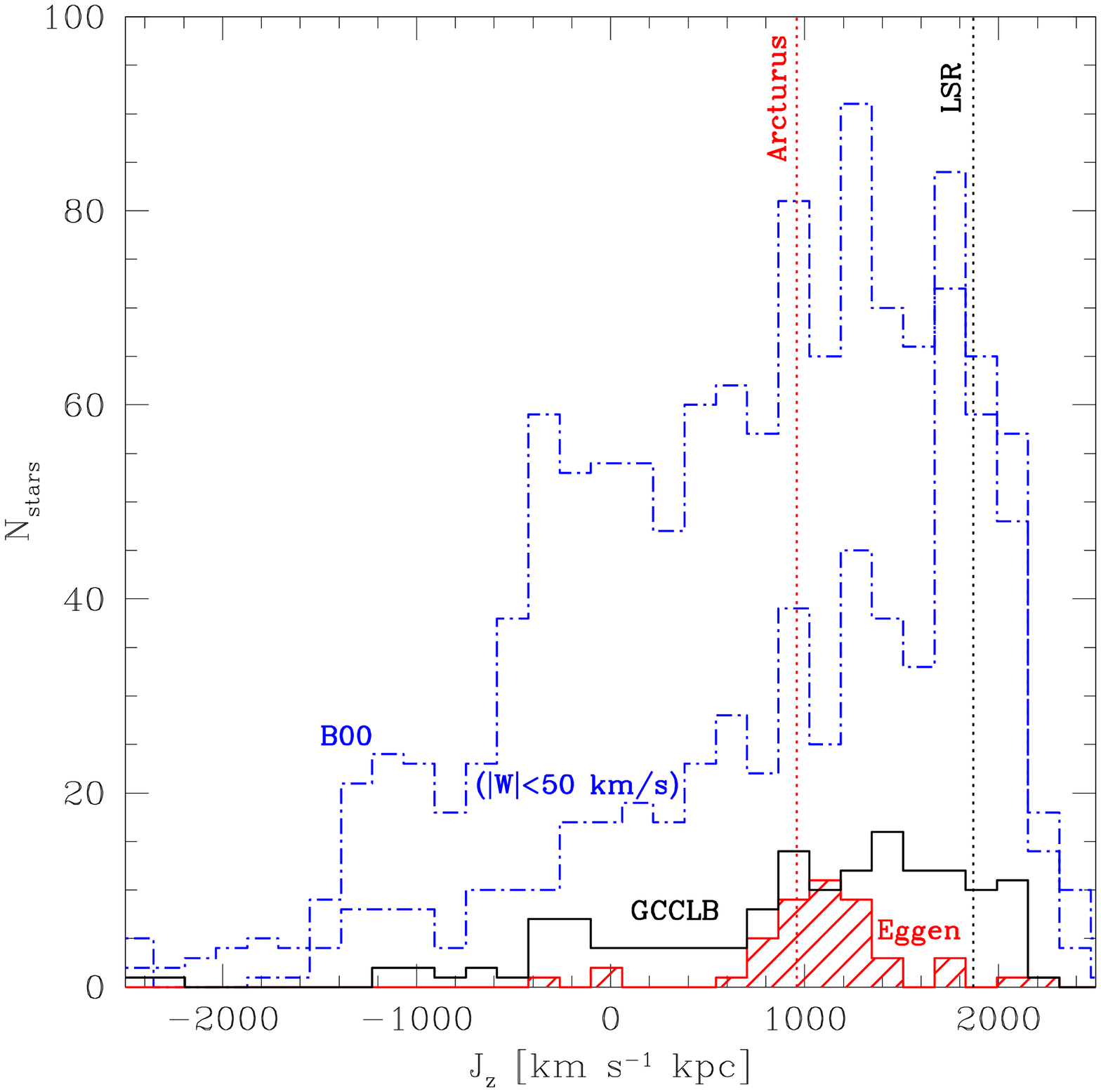}
\includegraphics[height=0.32\textheight,clip]{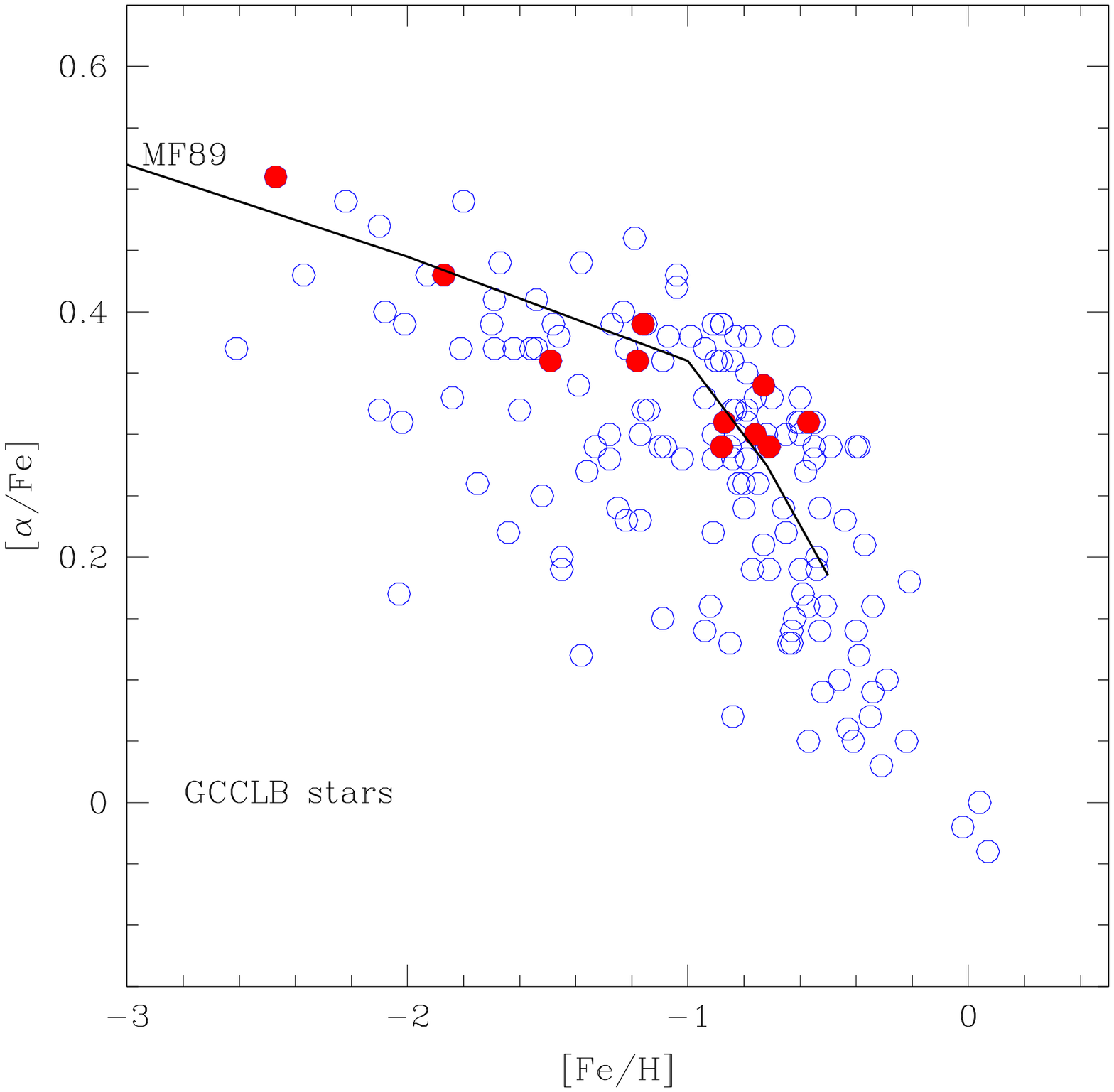}
\caption{ (a) Angular momentum distribution of stars in the Beers et
al (2000, B00) catalog (top two histograms); in the Gratton et al
(2003, GCCLB) compilation (solid histogram), as well as that of stars
identified as ``Arcturus group'' candidates by Eggen (1986, shaded
histogram). (b) $\alpha$-enhancement ([$\alpha$/Fe]) vs iron abundance
([Fe/H]) for all stars in the GCCLB compilation. Stars with angular
momentum similar to Arcturus are shown as solid circles, and compared
to a simple closed-box self-enrichment model (Matteucci \& Francois 1989).}
\end{figure*}


I wish to thank my collaborators, in particular Amina Helmi, Mario
Abadi, Andres Meza, and Matthias Steinmetz, for allowing me to
reproduce results of our collaborative program here. 

\end{document}